# Single-crystal growth and magnetic, magnetoelectric, and optical properties of ferroaxial-type $SrMn_2Ni_6Te_3O_{18}$

Ryoya Nakamura,[1] Shinichiro Asai,[2] Yusuke Nambu,[3] Takatsugu Masuda,[2,4] Kenta Kimura[1,5,6†]

[1]Department of Materials Science, Osaka Metropolitan University, Osaka 599-8531, Japan
[2]Institute for Solid State Physics, The University of Tokyo, Chiba 277-8581, Japan
[3]Institute for Integrated Radiation and Nuclear Science, Kyoto University, Kumatori 590-0494, Japan
[4]Institute of Materials Structure Science, High Energy Accelerator Research Organization, Tsukuba 305-0801, Japan
[5]Center for Spintronics Research Network (CSRN), Graduate School of Engineering Science, The University of Osaka, Toyonaka, Osaka 560-8531, Japan
[6]Innovative Quantum Material Center (IQMC), Osaka Metropolitan University, Osaka 558-8585, Japan

[†]Corresponding author: kentakimura@omu.ac.jp

## Abstract

Single crystals of $SrMn_2Ni_6Te_3O_{18}$, a member of the ferroaxial-type magnetic oxide family $AB_2C_6Te_3O_{18}$ ($A$ = Pb, Sr; $B$ = Mn, Cd; $C$ = Ni, Co), have been successfully grown, and their structural, magnetic, magnetoelectric, and optical properties have been systematically studied. Imaging of the spatial distribution of electric-field-induced optical rotation reveals that the single crystals preferentially form single ferroaxial (FA) domains. Magnetization and neutron diffraction measurements show that $Mn^{2+}$ and $Ni^{2+}$ magnetic moments order antiferromagnetically at $T_N$ = 83 K, forming a $c$-axis collinear bidirector-type antiferromagnetic structure. All independent magnetoelectric tensor components allowed by the magnetic point group $6/m'$ have been detected, and the $\chi_{33}$ component exhibits a pronounced temperature-dependent anomaly, including a peak and a sign reversal. Preferential formation of single FA domains and a similar $\chi_{33}$ anomaly are also observed in the isostructural compound $PbMn_2Ni_6Te_3O_{18}$. These findings suggest that the ferroaxial and magnetic characteristics within this structural framework are robust against Sr-Pb replacement.

## I. INTRODUCTION.

Ferroic orders such as ferroelectricity and ferromagnetism, which arise from the alignment of electric or magnetic dipoles, play key roles in modern technologies [1,2]. These ferroic orders originate from the breaking of specific material's symmetries: ferroelectricity requires the loss of spatial inversion ($\mathcal{P}$) symmetry while ferromagnetism requires the breaking of time-reversal ($\mathcal{T}$) symmetry [3]. In addition to these conventional ferroics, magnetic order characterized by vortex-like arrangements of magnetic dipoles has recently attracted considerable attention [4–7]. Such ferrotoroidal (FT) orders break $\mathcal{P}$ and $\mathcal{T}$ symmetries and can induce unconventional electromagnetic responses, including the linear magnetoelectric (ME) effect, in which electric polarization is induced by a magnetic field or magnetization is induced by an electric field [8], nonreciprocal propagation of (quasi-)particles [9–15], and current-induced magnetization [16]. Although various materials exhibiting FT order have been reported, the underlying symmetry mechanisms and material realizations continue to be actively explored [6,16–21].

Another ferroic class relevant to symmetry-dependent electromagnetic responses is the ferroaxial (FA) order, which arises from rotational distortion of structural units or vortex-like arrangements of electric dipoles [22–26]. FA order preserves $\mathcal{P}$ and $\mathcal{T}$ symmetries, but breaks rotational symmetry perpendicular to the axial direction (i.e., the vortex plane). When combined with certain helical magnetic structures, FA order has been reported to induce electric polarization [22]. However, there is a limited number of materials with well characterized FA order and magnetic order, and the correlation between FA order and magnetic order remains insufficiently understood.

The $AB_2C_6Te_3O_{18}$ family of oxides ($A$ = Pb, Sr; $B$ = Mn, Cd; $C$ = Ni, Co) crystallizes into the centrosymmetric hexagonal structure with space group $P6_3/m$ (point group $6/m$), as shown in Fig. 1(a) [21,27–29]. This group allows the FA-type structural distortion about the $c$ axis [21,23,24]. Due to the complexity of this structure, the FA nature is more conveniently described by rotational distortion of structural units than by local electric dipoles. Specifically, Figs. 1(b) and 1(c) visualize the FA nature as the unidirectional rotational arrangement of $CO_6$ octahedra, in which clockwise [Fig. 1(b)] and counterclockwise [Fig. 1(c)] rotations correspond to a pair of energetically equivalent FA domain states. Because this framework can accommodate various magnetic ions, it offers an advantageous platform for systematically studying how the interplay between FA structural distortion and magnetism influences electromagnetic responses. Among this family, $SrMn_2Ni_6Te_3O_{18}$ (hereafter abbreviated as SrMNTO) was first synthesized in the powder form and structurally characterized in 2020 [29]. The initial report described its coordination geometry, lattice parameters, and optical characteristics in detail. However, despite the presence of magnetic $Mn^{2+}$ and $Ni^{2+}$ ions, its magnetic properties had not been investigated, and the interplay between FA-type structural distortion and magnetism remained completely unresolved.

In the present study, we successfully synthesized single crystals of SrMNTO and carried out comprehensive measurements of their magnetic, dielectric, ME, and nonreciprocal optical responses. The magnetic structure was also examined using neutron powder diffraction. FA-type materials can adopt the two energetically equivalent FA domain states, as illustrated in Figs. 1(b) and 1(c), and understanding their spatial distribution is essential to characterize the FA order. To address this, we examined the FA domains in single crystals using a linear electrogyration (EG) effect, an optical phenomenon in which optical rotation is induced linearly by an applied electric field ($E$) [30]. Because applying an electric field along the $c$ axis breaks the mirror symmetry and lowers the point group symmetry from $6/m$ to 6, the system can acquire chirality and exhibit optical rotation. Since the sign of the induced optical rotation reflects the FA domain state, EG measurements provide a probe of the FA domain distribution [26]. For reference, we also performed limited measurements on the isostructural compound $PbMn_2Ni_6Te_3O_{18}$ (abbreviated as PbMNTO), primarily to compare FA domains and ME responses.

## II. METHODS

Polycrystalline samples of SrMNTO were synthesized by a solid-state reaction following the procedure described in Ref. [29]. A mixture of stoichiometric amounts of $SrCO_3$, $MnCO_3$, NiO, and $TeO_2$ powders was calcined at 1123 K in air for 12 hours. This process was repeated several times to ensure phase purity. Single crystals of SrMNTO were grown using a $TeO_2$ self-flux method. A mixture of $SrCO_3$, $MnCO_3$, NiO, and $TeO_2$ powders with a molar ratio of Sr:Mn:Ni:Te = 1:3:5:5 was placed in a Pt crucible, heated to 1323 K and kept for 12 hours, followed by slow cooling to 1223 K at a rate of 1 K/h before cooling to room temperature. Phase identification was performed by powder X-ray diffraction (XRD) using an X-ray diffractometer (Ultima IV, Rigaku). The crystallographic orientation of the single crystals was determined using the $\omega$-scan function of the same instrument.

The spatial distribution of FA domains was examined at room temperature by a recently developed FA-domain imaging technique that combines polarization microscopy with a field-modulation imaging method [26,31]. A plate-shaped single crystal specimen, whose largest face was parallel to the $c$ plane and whose thickness was approximately 0.1 mm, was placed between a polarizer and an analyzer that were set at relative angle of $\theta = +45°$ or $-45°$. Transparent ITO films were sputtered onto both faces of the sample, and electrical contacts were made using silver paste at the film edges to apply a voltage. The $c$ axis of the specimen was aligned with the direction of the incident light with a wavelength of 780 nm. Under these conditions, transmission microscope images were recorded using a scientific CMOS camera (Quantalux sCMOS, Thorlabs), while a rectangular-wave voltage was applied along the $c$ axis with stepwise ramps to suppress abrupt voltage changes. The quantity $\Delta I/I = \frac{I_+ - I_-}{(I_+ + I_-)/2}$ was then calculated at each camera pixel, where $I_+$ and $I_-$ denote the image intensities acquired at the positive and negative peak voltages $\pm V$, respectively. To reduce noise, $\Delta I/I$ was averaged over 20,000 images, and a 3 × 3 median filter was applied. For SrMNTO, which has the centrosymmetric FA structure with point group $6/m$, natural optical rotation is absent. Since EG-induced rotation angle $\varphi$ is typically small ($\sin\varphi \ll 1$) [26,30], $\Delta I/I$ at $\theta = \pm 45°$ can be expressed as

$$\Delta I/I = \pm\frac{4\pi}{180}\varphi = \pm\frac{4\pi}{180}\beta V,$$

where $\beta$ is the coefficient representing the magnitude of the linear EG effect, whose sign reflects the sign of the FA order parameter. Therefore, the spatial distribution of $\Delta I/I$ directly corresponds to the FA domain distribution.

Magnetic properties were characterized by a magnetic property measurement system (MPMS3, Quantum Design). The temperature ($T$) dependence of magnetization ($M$) was measured under an external magnetic field ($H$) of 1 kOe applied both parallel and perpendicular to the $c$ axis. For the evaluation of ME properties, the pyroelectric or $H$-induced current along the $[1\bar{1}0]$ and [001] directions were measured under $H$ up to 60 kOe using an electrometer (6517B, Keithley). The sample was cooled from the paramagnetic phase across $T_N$ under applied cooling electric ($E_c$) and cooling magnetic ($H_c$) fields to prepare an antiferromagnetic single-domain state, a procedure referred to as ME cooling. After ME cooling, $E_c$ was switched off. Electric polarization ($P$) was then obtained by integrating the current over time. Electrodes were formed using silver paste. $T$ and $H$ were controlled using a physical property measurement system (PPMS, Quantum Design).

Optical absorption coefficient ($\alpha$) spectra for unpolarized light propagating along the $c$ axis in the photon-energy range $1.3 < E_{ph} < 2.5$ eV were measured on a 55-μm-thick sample with a polished (001) surface using a home-built fiber-based optical setup [15]. White light from a tungsten-halogen lamp (AvaLight-HAL-S-MINI, Avantes) was delivered to the sample through an optical fiber, and the transmitted light was collected with another fiber and directed to a multichannel spectrometer (Flame-S, Ocean Insight) with an optical resolution of 1.5 nm. This setup is suitable for investigating nonreciprocal directional dichroism (NDD) because the direction of light propagation ($\boldsymbol{k}$) can be reversed ($\boldsymbol{k}\|+c \leftrightarrow \boldsymbol{k}\|-c$) simply by swapping the optical fibers connected to the light source and the spectrometer. The measurement system was integrated with a PPMS to control $T$ and apply $H_c$. Prior to each measurement, ME cooling was applied, after which $E_c$ and $H_c$ were switched off to measure the spontaneous NDD signal. $E_c$ and $H_c$ were parallel to the $[1\bar{1}0]$ and [110] directions, respectively. To apply $E_c$, a pair of parallel electrodes with a separation of 1 mm was formed on one surface of the sample using silver paste.

To clarify the magnetic structure, neutron powder diffraction (NPD) experiments were conducted using approximately 8 g of polycrystalline samples by means of the HERMES diffractometer [32] installed at JRR-3, employing an incident neutron wavelength of 2.19770(27) Å. The diffraction profiles were evaluated by carrying out Rietveld refinements with the FullProf software package [33]. Temperature control during the measurements was achieved using a 4 K closed-cycle helium refrigerator.

The crystal structures in Figs. 1(a)-1(c) and the magnetic structure in Fig. 6(b) were drawn using VESTA software [34].

## III. RESULTS and DISCUSSION

### A. Sample Characterization

The inset of Fig. 2 shows a photograph of a representative as-grown single crystal. Dark green, translucent single crystals with elongated tabular shapes and sizes up to a few millimeters have been successfully obtained using the self-flux method. The red curve in Fig. 2 presents the powder XRD pattern of pulverized samples. The pattern is in good agreement with the simulated pattern for SrMNTO, with no impurity phases detected, confirming the successful growth of single-phase SrMNTO crystals. The blue curve in Fig. 2 shows the XRD pattern obtained by irradiating x-rays onto the widest surface of the as-grown crystal. Only $h00$ reflections are observed, indicating that this surface is parallel to (100). The long axis of most as-grown crystals corresponds to the [001] direction.

### B. Ferroaxial Domains

Figure 3(a) displays a transmission optical micrograph of the single crystal used for the EG measurements. The small dark spots scattered across the sample arise from micropores formed during the sample polishing process. The large dark areas marked by arrows are regions with residual silver paste. Spatial maps of $\Delta I/I$ for the same region as in Fig. 3(a) were obtained under various applied voltages, reaching amplitudes up to 300 V, in polarization geometries with $\theta = \pm 45°$. The resulting images are shown in Figs. 3(b)-3(i), where red and blue correspond to positive and negative values of $\Delta I/I$, respectively. As $V$ increases, the entire area gradually shifts to red at $\theta = 45°$, whereas it becomes blue at $-45°$. This reversal of sign is a hallmark of electric-field-induced optical rotation, i.e., EG, as described by Eq. (1).

To determine whether the observed EG originates from the linear contribution or includes higher-order effects, we evaluated the averaged $\Delta I/I$ over selected regions marked by boxes in Figs. 3(b)-(i) and plotted the values as a function of applied $V$. As shown in Fig. 3(j), the magnitude of $\Delta I/I$, which corresponds to the EG magnitude, increases in proportion to $V$. From the slope, the linear EG coefficient $\alpha$ was evaluated to be $2 \times 10^{-2}$ deg/kV, which is typical value [26,30]. These results demonstrate that $\Delta I/I$ reflects the linear EG response and therefore indicate that the investigated sample contains a single FA domain. A similar single FA domain was also observed in another crystal, suggesting that a single FA domain is preferred in SrMNTO. It also indicates that SrMNTO does not undergo a structural phase transition to a non-ferroaxial phase below the crystallization temperature. Nucleation and crystal growth proceed in the FA phase without forming additional domains because domain formation requires domain wall energy and is not energetically favorable. For reference, we also examined the FA domain distribution in a single crystal of the isostructural compound PbMNTO, and found that the crystal likewise exhibits a single FA domain (see Fig. S1 of the Supplemental Material [35]). Comparable single FA domains have been reported in other FA materials that do not experience a high-temperature FA transition, including $PbWO_4$ [26] and $MnTiO_3$ [36].

### C. Magnetic Properties

Figure 4(a) shows the $T$ dependence of magnetic susceptibility (defined as $\chi_m = M/H$), where $\chi_{m\parallel c}$ and $\chi_{m\perp c}$ are the susceptibilities measured under $H$ = 1 kOe applied parallel and perpendicular to the $c$ axis, respectively. At high temperatures, the susceptibilities are essentially isotropic. Above 150 K, both $\chi_{m\parallel c}$ and $\chi_{m\perp c}$ follow the Curie-Weiss law, $\chi_m = \chi_0 + C/(T - \theta_{CW})$, as shown in the inset. Here, $\chi_0$ is the constant term, $C$ is the Curie constant, and $\theta_{CW}$ is the Weiss temperature. Fitting yields an effective moment per formula unit (containing two $Mn^{2+}$ and six $Ni^{2+}$ ions) of $\mu_{eff}$ = 11.62 $\mu_B$/f.u. and 11.76 $\mu_B$/f.u. for $\chi_{m\parallel c}$ and $\chi_{m\perp c}$, respectively. Theoretically, assuming complete quenching of the orbital angular momentum, $\mu_{eff}$ obtained from the Curie constant is expressed as $\sqrt{2g_{Mn}^2 S_{Mn}(S_{Mn}+1)+6g_{Ni}^2 S_{Ni}(S_{Ni}+1)}\mu_B$, where $S_{Mn}$ = 5/2 and $g_{Mn}$ = 2 for $Mn^{2+}$, and $S_{Ni}$ = 1 and $g_{Ni}$ = 2 for $Ni^{2+}$, giving 10.86 $\mu_B$/f.u. This value is only about 10% smaller than the experimental ones, indicating that the orbital angular moment is almost quenched and contributes only through a slight $g$-factor modification via spin-orbit coupling. The derived $\theta_{CW}$ values are −112.4 K and −113.4 K for $\chi_{m\parallel c}$ and $\chi_{m\perp c}$, respectively, confirming that antiferromagnetic interactions dominate.

Upon cooling, both $\chi_{m\parallel c}$ and $\chi_{m\perp c}$ show clear anomalies at $T_N$ = 83 K, indicating the onset of an antiferromagnetic transition. Below $T_N$, $\chi_{m\parallel c}$ and $\chi_{m\perp c}$ exhibit markedly different behaviors: $\chi_{m\parallel c}$ decreases sharply, whereas $\chi_{m\perp c}$ remains nearly constant. This anisotropic behavior below $T_N$ is characteristic of a uniaxial antiferromagnet with magnetic moments aligned along the $c$ axis. The upturn observed at low temperatures below 10 K is likely due to a small Curie-tail contribution from paramagnetic defects or trace impurities whose concentrations are below the detection limits of X-ray and neutron powder diffraction (as described later).

**D. Neutron Diffraction**

Figure 5(a) displays the NPD patterns recorded at various temperatures between 5 and 300 K. All diffraction peaks at 300 K can be indexed using the $P6_3/m$ space group, in agreement with the X-ray diffraction results. The peaks observed at 5 K are also indexed with the same space group; however, several reflections show enhanced intensities compared with those at 300 K. To clarify the origin of this enhancement, the integrated intensity of 011 reflection is plotted as a function of temperature [Fig. 5(b)]. The intensity increases below $T_N$, demonstrating that the additional contribution arises from magnetic scattering.

The magnetic reflections occur at the same positions as the nuclear reflections within experimental resolution, indicating that the magnetic unit cell coincides with the crystallographic unit cell and is characterized by the magnetic propagation vector $\mathbf{q} = (0, 0, 0)$. No $00l$ magnetic reflections are observed, implying that the ordered moments are aligned parallel to the $c$ axis, consistent with the anisotropic behavior of $\chi_m$ [Fig. 4(a)]. To determine the magnetic structure through Rietveld refinement, representation analysis was carried out for the space group $P6_3/m$ with $\mathbf{q} = (0, 0, 0)$ using the SARA*h* code [37]. The magnetic representations for $Ni^{2+}$ at the $12i$ site and $Mn^{2+}$ at the $4f$ site decompose as $\Gamma_{Ni} = 3\sum_{i=1}^{12}\Gamma_i^{(1)}$ and $\Gamma_{Mn} = \sum_{i=1}^{12}\Gamma_i^{(1)}$, respectively, as listed in Tables S1 and S2 of the Supplemental Material [35]. Among these, the $\Gamma_2$, $\Gamma_7$, and $\Gamma_8$ models are compatible with a $c$-axis collinear antiferromagnetic arrangement. For $Ni^{2+}$, three basis vectors ($\Psi_1$, $\Psi_2$, and $\Psi_3$) are allowed, and only $\Psi_3$, corresponding to the $c$-axis component, was retained to enforce the collinear structure. In contrast, $Mn^{2+}$ has a single basis vector ($\Psi_1$) corresponding to the $c$-axis component. Figure S2 of the Supplemental Material [35] shows simulated magnetic diffraction profiles for the $\Gamma_2$, $\Gamma_7$, and $\Gamma_8$ models, where the $c$-axis component ($Ni^{2+}$: $\Psi_3$, $Mn^{2+}$: $\Psi_1$) were refined. Only the $\Gamma_2$ model provides a satisfactory description of the measured data. Therefore, the 5 K data was refined using the $\Gamma_2$ model. In the refinement, isotropic atomic displacement parameters were constrained to be identical for the Sr, Ni, Mn, and Te sites, and likewise identical among the four O sites.

Figure 6(a) presents the refined pattern together with the reliability factors, showing good agreement with the experiment. The corresponding magnetic structure is illustrated in Fig. 6(b). The unit cell contains two Mn-Mn pairs and six Ni-Ni pairs arranged along the $c$ axis, where the two ions in each pair are related by a mirror operation (orange plane). The magnetic moments within each pair are antiparallel and aligned along the line connecting the two ions, and thus the magnetic structure corresponds to a bidirector-type (or falsely chiral [23]) antiferromagnetic configuration. This magnetic structure is identical to that of the isostructural compound PbMNTO and belongs to the magnetic point group $6/m'$ [21,28]. It breaks both $\mathcal{P}$ and $\mathcal{T}$ symmetries, allowing the linear ME effect formulated as $P = \chi H$ and $M = {}^t\chi E$. Here, $\chi$ represents the ME tensor, whose nonzero components are symmetric diagonal components ($\chi_{11} = \chi_{22}$ and $\chi_{33}$) as well as antisymmetric off-diagonal components ($\chi_{12} = -\chi_{21}$); all other tensor components vanish. The antisymmetric components are characteristic of a ferrotoroidic state, which, in both SrMNTO and PbMNTO, is induced by a unique combination of FA-type structural distortions and bidirector-type antiferromagnetic order. The refined crystallographic and magnetic parameters are summarized in Table S3 of the Supplemental Material [35]. The ordered magnetic moments of $Ni^{2+}$ and $Mn^{2+}$ at 5 K are $2.04(2)\mu_B$ and $4.49(3)\mu_B$, respectively, which fall within the typical ordered-moment ranges expected for $Ni^{2+}$ and $Mn^{2+}$ in oxides.

Rietveld refinements were also performed for the datasets at other temperatures to examine the temperature evolution of the crystal and magnetic structures. Magnetic refinements were conducted only for temperatures at or below 80 K, because magnetic contributions to the diffraction intensities are not detectable above this temperature [see Fig. 5(b)]. The refined patterns are presented in Figures S3–S7 of the Supplemental Material [35]. Figure 7(a) shows the *T* dependence of the lattice constants *a* and *c*. A clear anomaly appears in the *c*-axis parameter at $T_N$, suggesting that the magnetic ordering is strongly coupled to the lattice along the *c* axis. Since SrMNTO contains edge-sharing zigzag chains of $NiO_6$ octahedra along the *c* axis (Fig. 6b), exchange interactions associated with the chain direction might be responsible for the observed magnetoelastic response. Figure 7(b) presents the *T* dependences of the ordered moments of Ni and Mn, plotted on the left and right axes, respectively. The two moments exhibit nearly identical *T* dependences, indicating that the two sublattices are closely coupled and are governed by a single magnetic order parameter.

**E. Magnetoelectric Properties**

The linear ME effect expected in the antiferromagnetic phase was examined by measuring the *H*-induced *P* (i.e., $P = \chi H$). For convenience, the [110] $[1\bar{1}0]$ and [001] axes are defined as *X*, *Y*, *Z*, respectively, and *P* (*H*) along the *i* axis ($i = X, Y, Z$) is denoted as $P_i$ ($H_i$). We measured $P_Y$ under $H_Y$, $P_Y$ under $H_X$, and $P_Z$ under $H_Z$, which can probe the three independent ME components, $\chi_{11}$, $\chi_{12}$, and $\chi_{33}$, respectively. The *T* dependences of the corresponding *P* at $H = 60$ kOe, measured after the ME cooling with $E_c = \pm 770$ kV/m and $H_c = 60$ kOe, are shown in Figs. 8(a)-8(c). *P* appears below $T_N$ for all the geometries, indicating nonzero $\chi_{11}$, $\chi_{12}$, $\chi_{33}$ components. The reversal of *P* between positive and negative $E_c$ for all the geometries shows that the antiferromagnetic domains can be controlled by the sign of $E_c$ during the ME cooling process. The *H* dependences of *P* after the ME cooling process with positive $E_c$ are shown in Figs. 8(d)-8(f). The linear relationship between *P* and *H* is clearly observed, demonstrating the linear ME effect.

As shown in Figs. 8(d)-8(f), upon cooling, the $H_Y$- and $H_X$-induced $P_Y$ values (equivalently, $\chi_{11}$ and $\chi_{12}$) increase monotonically, whereas the $H_Z$-induced $P_Z$ (equivalently $\chi_{33}$) exhibits a nonmonotonic *T* dependence, featuring a peak around 75 K and a sign reversal at 50 K. This anomalous *T* dependence of $\chi_{33}$ is further confirmed by comparing the $P_Z$-$H_Z$ slopes at 10, 40, and 70 K shown in Fig. 8(f). A similar peak and sign reversal in the $\chi_{33}$ component have been reported for the *c*-axis collinear antiferromagnet $Cr_2O_3$, a prototypical linear ME material, when both the field and the response are parallel to the direction of the ordered moments (*c* axis) [38]. Mostovoy and co-workers [39] theoretically demonstrated that a Heisenberg exchange-driven mechanism combined with spin fluctuations can give rise to a finite $\chi_{33}$ in $Cr_2O_3$ (often denoted as $\chi_{\parallel}$ in the literature), where $\chi_{33}$ is proportional to the product of the magnetic order parameter (ordered moment) and the susceptibility perpendicular to the *c* axis. This accounts for the peak observed at elevated temperatures in $Cr_2O_3$. Considering the similar *c*-axis collinear antiferromagnetic structure, we expect that a similar mechanism may also be at play in SrMNTO and explain the peak observed at 75 K. The origin of the sign reversal of $\chi_{33}$ in $Cr_2O_3$ has been discussed but remained unclear [38,40]. Since the present compound SrMNTO contains two different magnetic ions ($Ni^{2+}$ and $Mn^{2+}$), the situation is more complex than in $Cr_2O_3$. Therefore, we leave the origin of the sign reversal in SrMNTO for future study. For reference, we also examined the *T*-dependence of $\chi_{33}$ in the isostructural compound PbMNTO. As shown in Fig. S7 of the Supplemental Material [35], the data likewise exhibit a peak and a sign reversal in $\chi_{33}$, indicating that this behavior is common to this pair of isostructural compounds, rather than being specific to SrMNTO.

**F. Nonreciprocal Directional Dichroism**

As mentioned above, the linear ME effect measurements have demonstrated the presence of the finite off-diagonal ME component $\chi_{12}$ in SrMNTO. However, this measurement cannot distinguish whether this component is symmetric ($\chi_{12} = \chi_{21}$) or antisymmetric ($\chi_{12} = -\chi_{21}$). The NDD for unpolarized light propagating along the $c$ axis arises from the latter component and therefore serves as a suitable probe for the distinguishment. It is well known that the NDD can be detected as the difference between the absorption coefficient ($\alpha$) for one antiferromagnetic domain ($\alpha_+$) and another one ($\alpha_-$), $\Delta\alpha = \alpha_+ - \alpha_-$, for a fixed light propagation direction, and these opposite domains can be selected by reversing either $E_c$ or $H_c$ during the ME cooling. We measured the spectra at 75 K in the antiferromagnetic phase after the ME cooling with ($+E_c$,$+H_c$) and ($-E_c$,$+H_c$), which are referred to as $\alpha_+$ and $\alpha_-$, respectively, and subsequently switched off these fields. The results for the light propagation direction $\boldsymbol{k}||+c$ are presented in Fig. 9(a). A broad absorption band above 1.5 eV is probably due to spin-allowed $d$-$d$ transitions of $Ni^{2+}$ ions [21]. While the two spectra nearly collapse on top with each other, a noticeable difference is observed at around 1.5 eV, which is more evident in the difference spectrum $\Delta\alpha = \alpha_+ - \alpha_-$, as shown in Fig. 9(b). Furthermore, the $\Delta\alpha$ spectrum is completely reversed upon switching between $\boldsymbol{k}||+c \leftrightarrow \boldsymbol{k}||-c$. This nonreciprocal behavior confirms that $\Delta\alpha$ arises from NDD, demonstrating the presence of the antisymmetric ME component and consequently the FT state in SrMNTO.

In certain materials, replacing Sr with Pb significantly alters magnetic exchange interactions and can even change the resulting magnetic symmetry, as illustrated by the case of the $Sr(TiO)Cu_4(PO_4)_4$ and $Pb(TiO)Cu_4(PO_4)_4$ system [41]. In contrast, the present study finds no such differences between SrMNTO and PbMNTO. Both compounds exhibit the same FT state arising from the combination of the FA-type structural rotation and the bidirector-type antiferromagnetic structure, and they show very similar ME behavior, including the sign reversal of $\chi_{33}$. We also note that both compounds host a single ferroaxial domain. These observations suggest that the magnetic and ferroaxial characteristics in the $AB_2C_6Te_3O_{18}$ family are robust against Sr-Pb replacement at the $A$ site.

**IV. CONCLUSIONS**

In this study, we have grown single crystals and investigated the structural, magnetic, magnetoelectric, and optical properties of $SrMn_2Ni_6Te_3O_{18}$ (SrMNTO), a member of the $AB_2C_6Te_3O_{18}$ family of oxides ($A$ = Pb, Sr; $B$ = Mn, Cd; $C$ = Ni, Co). Visualizing the distribution of electric-field-induced optical rotation has shown that single crystals preferentially host single ferroaxial (FA) domains. Neutron diffraction has revealed a collinear bidirector-type antiferromagnetic structure, giving rise to a ferrotoroidal (FT) state originating from the combination of the FA-type structural rotation and the magnetic order. This FT character has been directly confirmed through the observation of nonreciprocal directional dichroism in the antiferromagnetic phase. Magnetoelectric (ME) measurements have detected all three independent ME tensor components (the diagonal $\chi_{11} = \chi_{22}$ and $\chi_{33}$, and the antisymmetric off-diagonal $\chi_{12} = -\chi_{21}$) allowed by the magnetic point group $6/m'$ that breaks both $\mathcal{P}$ and $\mathcal{T}$ symmetries. Among them, $\chi_{33}$ displays a pronounced nonmonotonic temperature dependence, with a peak near 75 K and a sign reversal around 50 K. These behaviors resemble those reported for $Cr_2O_3$, and the peak may arise from a similar exchange-driven mechanism involving spin fluctuations.

We have also investigated the FA domains and the $\chi_{33}$ response of the isostructural compound $PbMn_2Ni_6Te_3O_{18}$ (PbMNTO), and found qualitatively identical behaviors as in SrMNTO. These results suggest that the magnetic and ferroaxial characteristics in the $AB_2C_6Te_3O_{18}$ framework are robust against Sr-Pb replacement at the *A* site. The preferential formation of the single FA domains in SrMNTO and PbMNTO further suggests that this structural family provides an excellent materials platform for exploring symmetry-dependent physical phenomena associated with a well-defined ferroaxial state. Extending such studies to other members of the $AB_2C_6Te_3O_{18}$ family (e.g., *B* = Cd, *C* = Co) will be an intriguing direction for future work.

**ACKNOWLEDGMENTS**

We would like to thank Dr. Rieko Ishii for enlightening discussions on crystal growth. K.K. acknowledges partial support from JSPS KAKENHI (Grant Nos. JP23K17663, JP24K00575, JP26H00677, JP26H02194) and the Iketani Science and Technology Foundation. Y.N. was partially supported by JSPS KAKENHI (Grant Nos. JP22H05145, JP24K00572, JP25K01489, JP26H02027) and JST FOREST (Grant No. JPMJFR202V).

The neutron diffraction experiment at JRR-3 was carried out under the general user program managed by the Institute for Solid State Physics, the University of Tokyo (Proposal No. 24601), and supported by Center of Neutron Science for Advanced Materials, Institute for Materials Research, Tohoku University (Proposal No. 202408-CNKXX-0040).

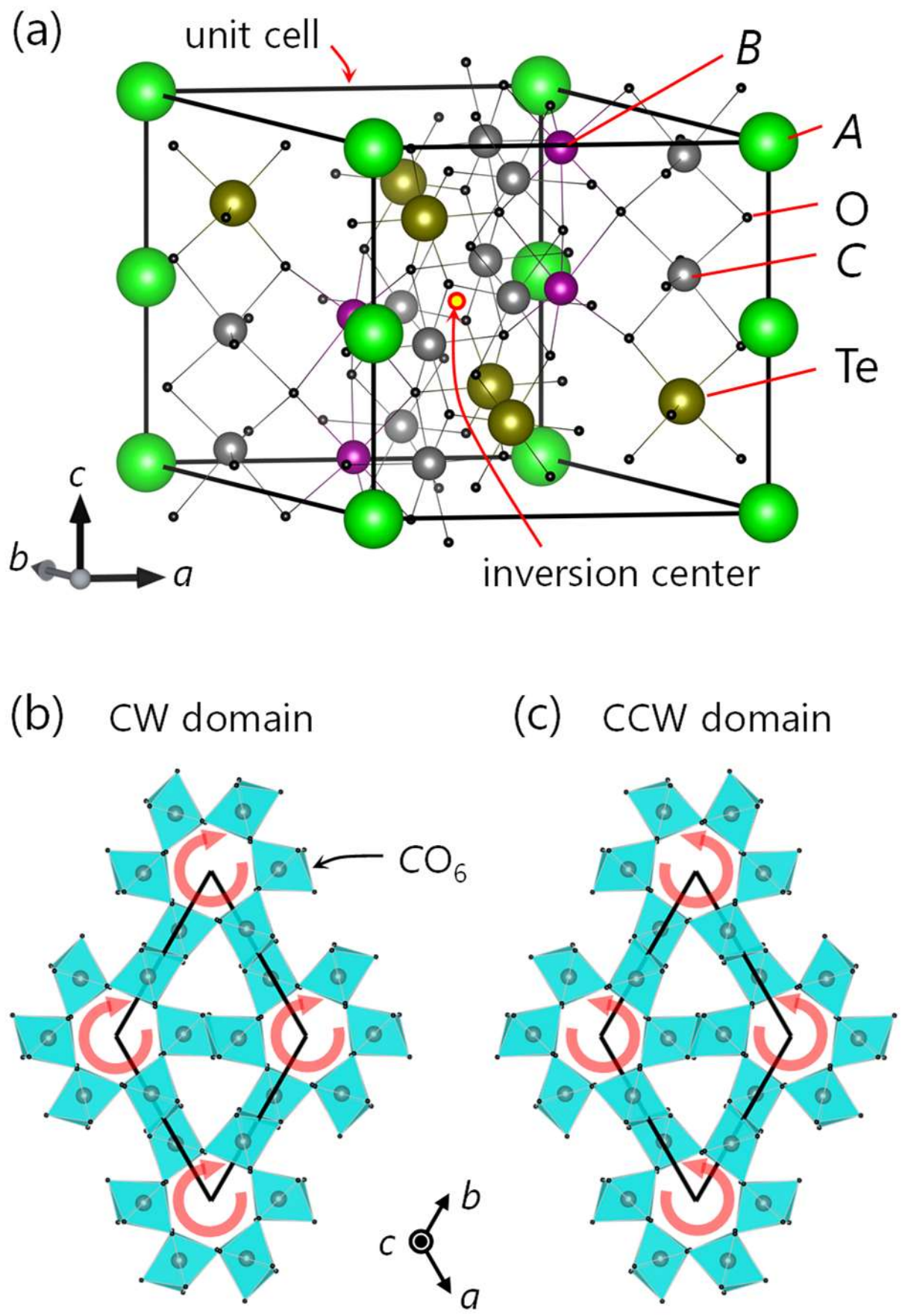

**Figure 1.** (Color online) (a) Crystal structure of the $AB_2C_6Te_3O_{18}$-type compound. Green, purple, gray, yellow, and black spheres represent *A*, *B*, *C*, Te, and O atoms, respectively. The black parallelogram denotes the unit cell. The red circle indicates a representative inversion center at the fractional coordinate (1/2, 1/2, 1/2). (b),(c) Projection of the crystal structure along the *c* axis, showing only $CO_6$ octahedra. Red curved arrows highlight clockwise (CW) (b) and counterclockwise (CCW) (c) rotational arrangements of the $CO_6$ octahedra, corresponding to a pair of energetically equivalent FA domain states.

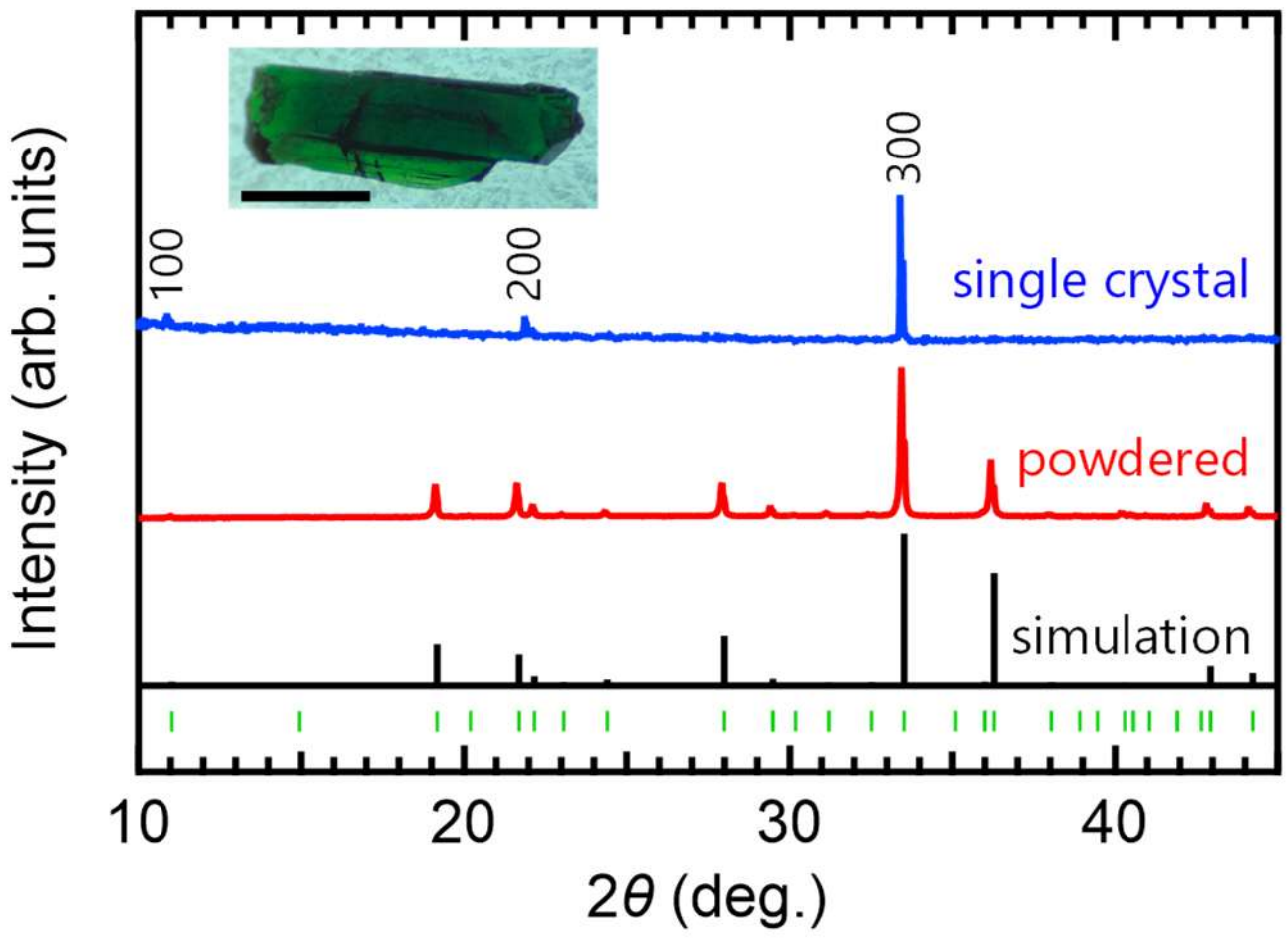

**Figure 2.** (Color online) X-ray diffraction profiles of single-crystal and powdered samples, along with the simulation. Green vertical bars indicate the positions of Bragg reflections. The inset shows a photo of the single crystal of SrMNTO. The scale bar is 1 mm.

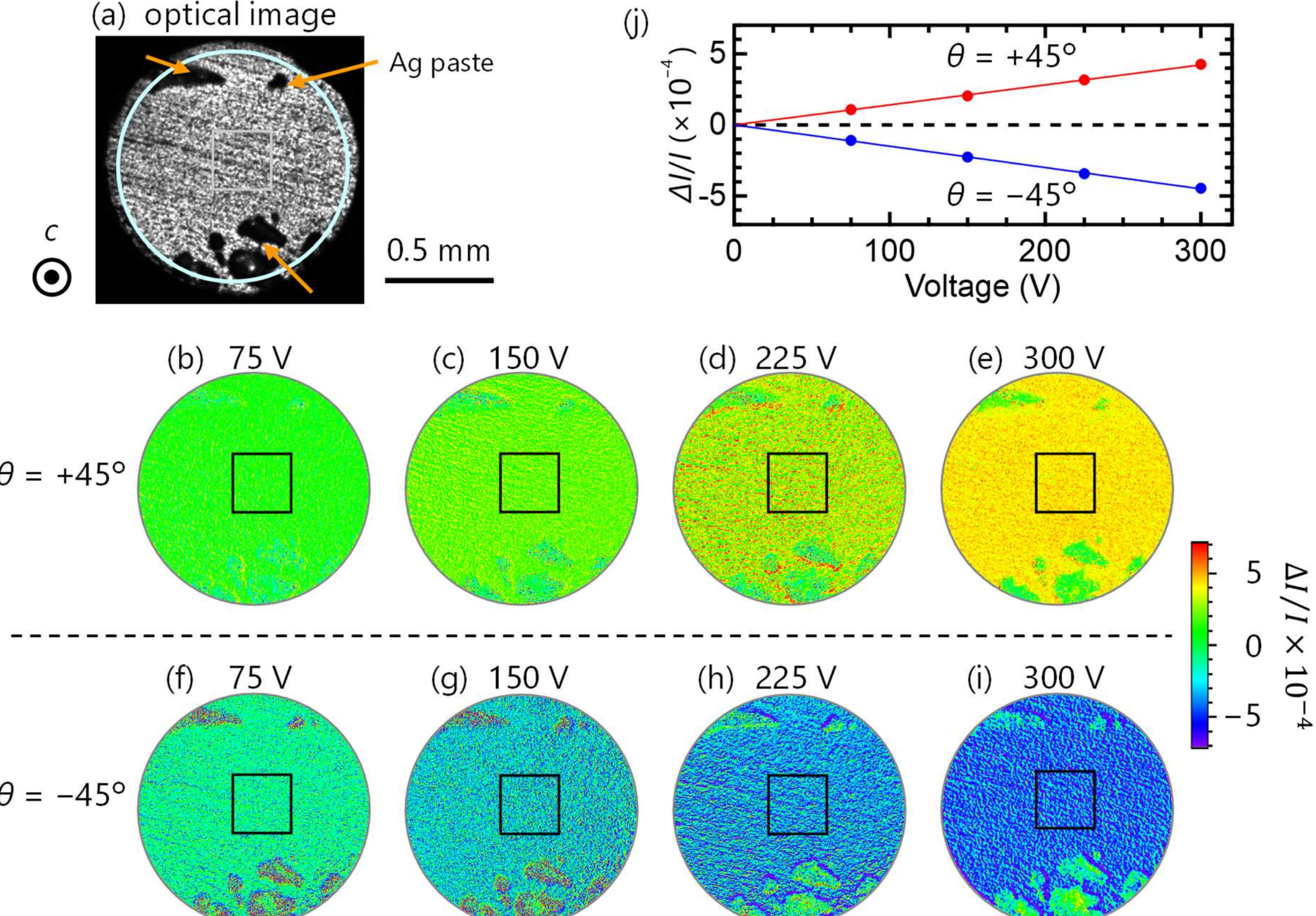

**Figure 3.** (Color online) Distribution of FA domains in a SrMNTO single crystal visualized by the EG effect. (a) Optical image. Orange arrows indicate Ag paste on the sample surface. (b)-(i) Spatial maps of $\Delta I/I$ at 75, 150, 225, and 300 V and at $\theta$ = ±45° for the region indicated by the circle in the panel (a). (j) Voltage dependence of $\Delta I/I$ averaged over the region marked by the black boxes in (b)-(i).

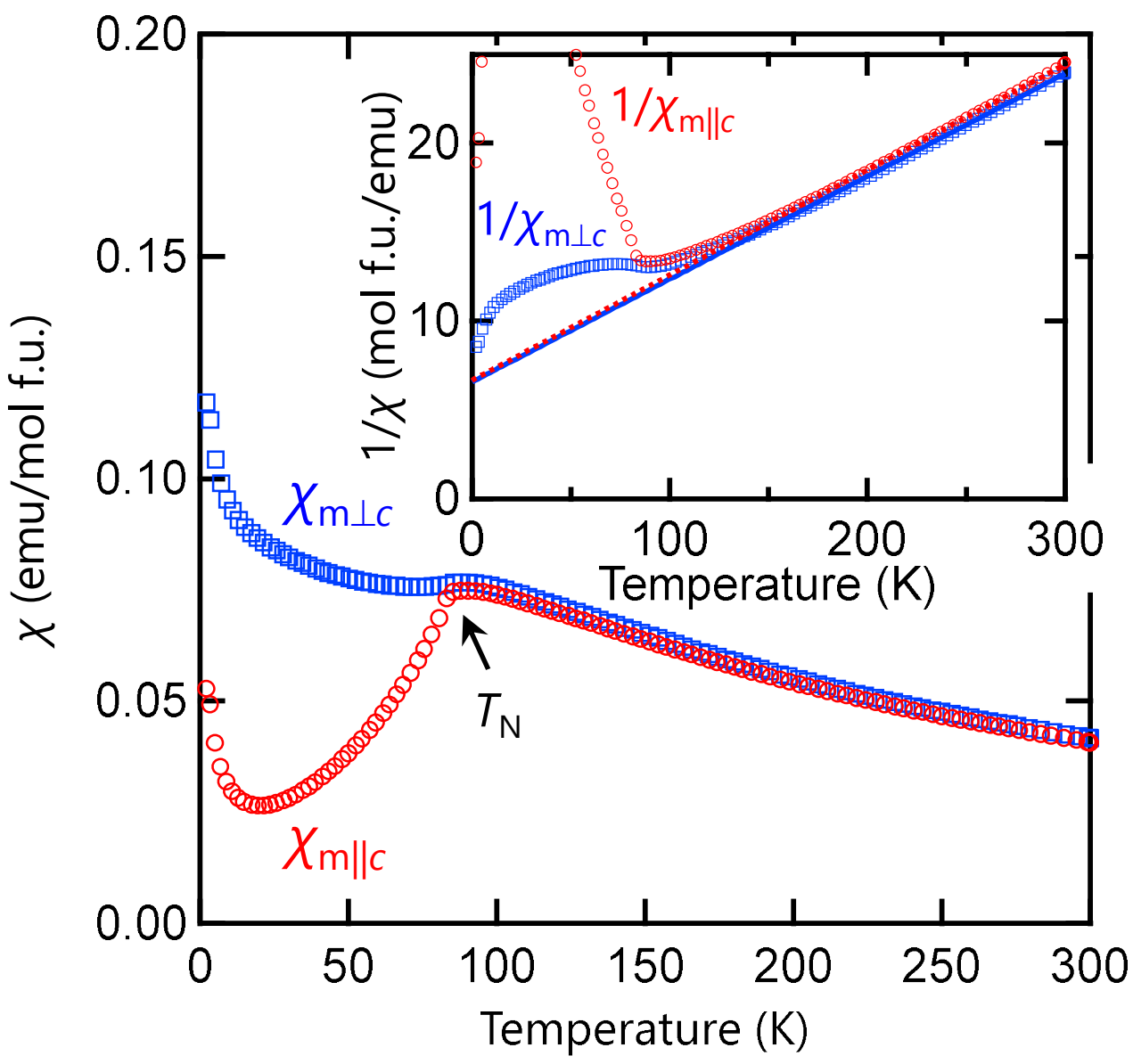

**Figure 4.** (Color online) Temperature dependence of the magnetic susceptibility of SrMNTO under an applied field of 1 kOe, parallel ($\chi_{\mathrm{m}\parallel c}$, red circles) and perpendicular ($\chi_{\mathrm{m}\perp c}$, blue squares) to the *c* axis. The inset shows the inverse susceptibility along with Curie-Weiss fits (red dotted line for $1/\chi_{\mathrm{m}\parallel c}$ and blue solid line for $1/\chi_{\mathrm{m}\perp c}$).

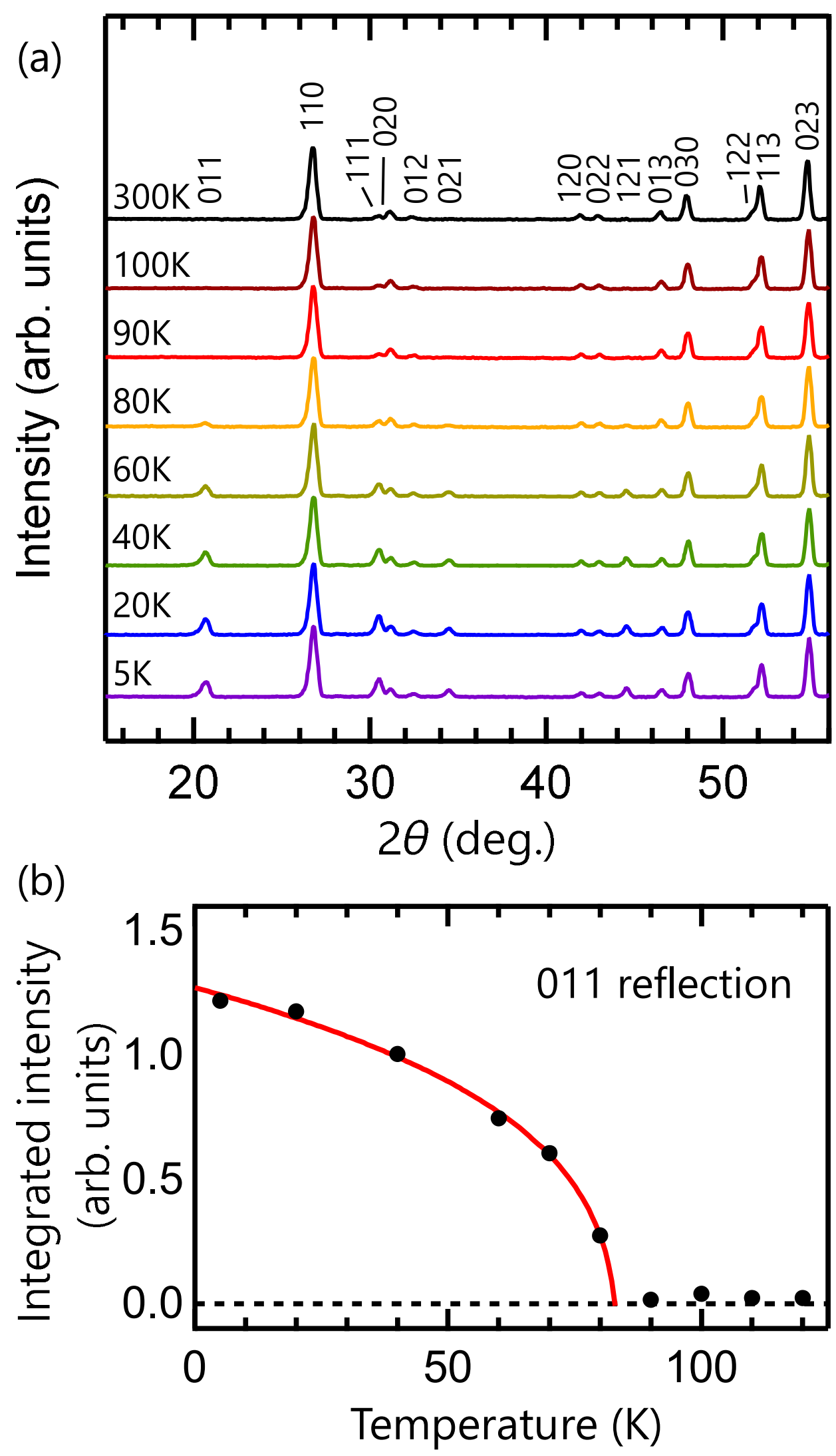

**Figure 5.** (Color online) (a) Neutron powder diffraction patterns of SrMNTO at various temperatures between 5 and 300 K. The main Bragg reflections are indexed with *hkl* indices. (b) Temperature dependence of the integrated intensity of 011 reflection. The red curve is the guide to the eye and was not used to determine the magnetic order parameter.

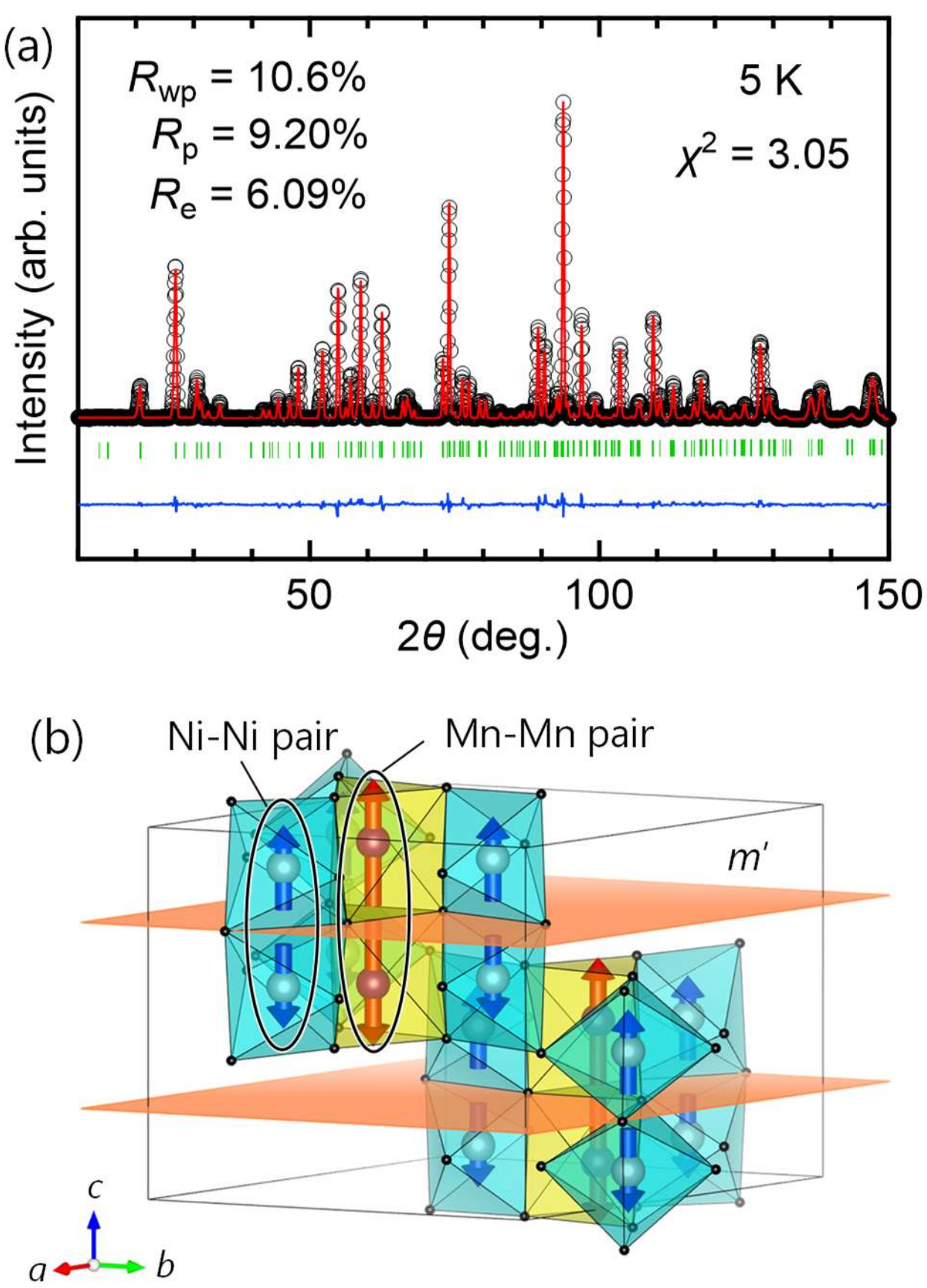

**Figure 6.** (Color online) (a) Observed (open circle), Rietveld refined (red curve), and difference (blue curve) neutron diffraction patterns of SrMNTO at 5 K. Green vertical bars denote the positions Bragg reflections. (b) Obtained magnetic structure. Gray, purple, and black spheres represent Ni, Mn, and O atoms, respectively. Blue and red arrows denote Ni and Mn magnetic moments, respectively, both oriented along the *c* axis. There are six Ni-Ni pairs and two Mn-Mn pairs in the unit cell (gray box), each of which is symmetric with respect to the mirror plane combined with time reversal (*m*′) perpendicular to the *c* axis (orange planes).

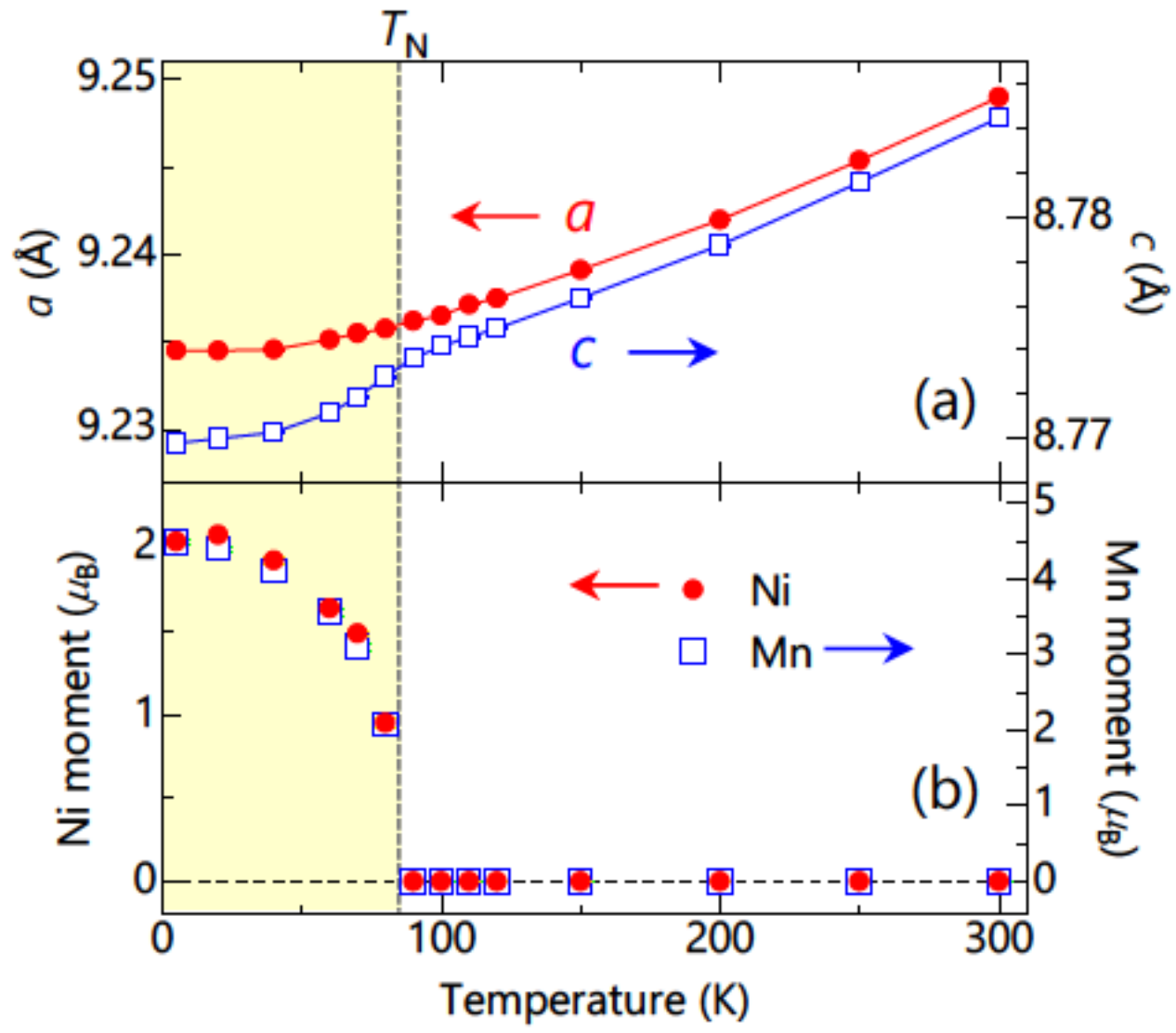

**Figure 7.** (Color online) (a) Temperature dependence of the lattice constants *a* (left axis) and *c* (right axis). (b) Temperature dependence of the magnetic moments of $Ni^{2+}$ (left axis) and $Mn^{2+}$ (right axis).

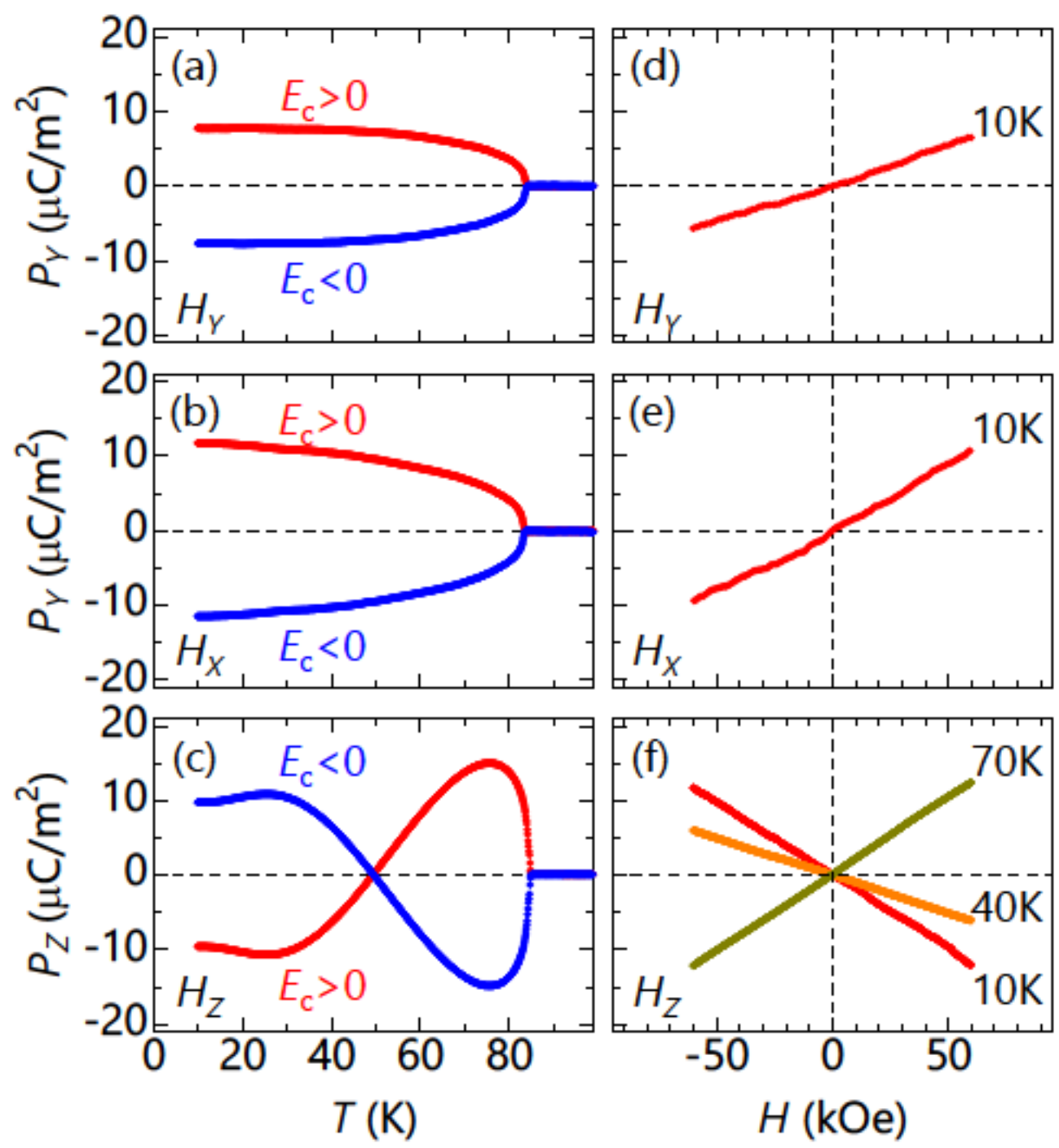

**Figure 8.** (Color online) (a-c) Temperature dependence of (a) $P_Y$ at $H_Y$ = 60 kOe, (b) $P_Y$ at $H_X$ = 60 kOe, and (c) $P_Z$ at $H_Z$ = 60 kOe for SrMNTO. These measurement geometries probe the $\chi_{11}$, $\chi_{12}$, and $\chi_{33}$ components, respectively. Red and blue symbols correspond to cooling electric fields of $E_c$ = +770 kV/m and −770 kV/m, respectively. (d) $P_Y$ vs. $H_Y$ at 10 K. (e) $P_Y$ vs. $H_X$ at 10 K. (f) $P_Z$ vs. $H_Z$ at 10, 40, and 70 K. For (d)-(f), $E_c$ = +770 kV/m.

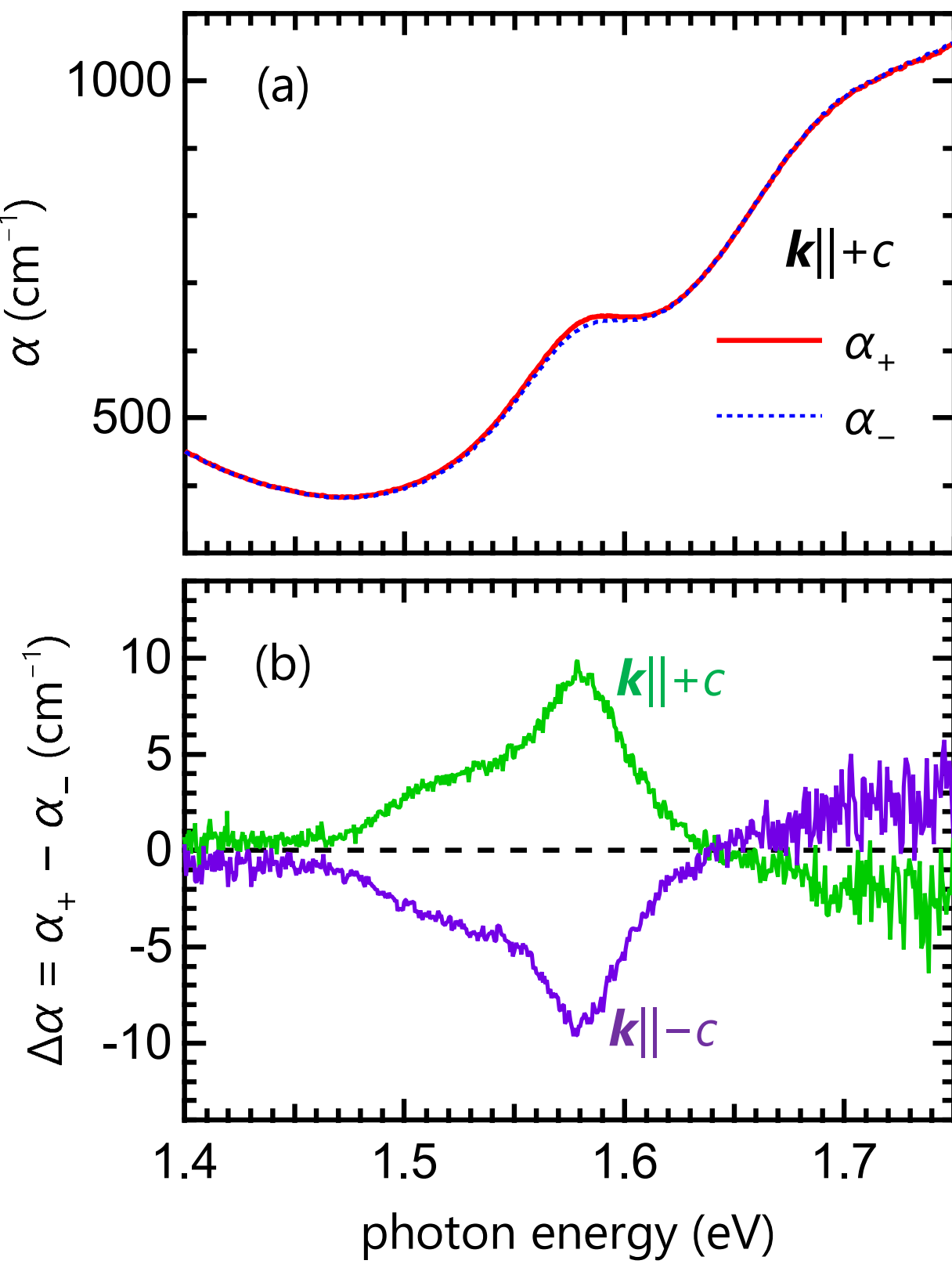

**Figure 9.** (Color online) (a) Absorption coefficient spectra of SrMNTO at 75 K for unpolarized light propagating along the $+c$ direction ($\boldsymbol{k}$||$+c$). The red solid and blue dotted curves correspond to $\alpha_+$ and $\alpha_-$, respectively, measured after ME cooling under ($+E_c$, $+H_c$) and ($-E_c$, $+H_c$). (b) Difference spectra ($\Delta\alpha = \alpha_+ - \alpha_-$) at 75 K for $\boldsymbol{k}$||$+c$ and $\boldsymbol{k}$||$-c$.